\documentclass[12pt,epsf]{article}

\usepackage{times}
\usepackage{graphicx}
\usepackage{amsfonts}
\usepackage{amsmath}
\usepackage{amssymb}
\usepackage{amstext}
\usepackage{amsthm}
\usepackage{mathtools}

\usepackage{epsfig}
\usepackage{color}
\psfigdriver{dvips}
\usepackage{latexsym}
\usepackage[matrix,curve]{xypic}
\usepackage{rotating}
\rotdriver{dvips}

\begin{document}

\baselineskip 6mm
\renewcommand{\thefootnote}{\fnsymbol{footnote}}


\newcommand{\nc}{\newcommand}
\newcommand{\rnc}{\renewcommand}


\rnc{\baselinestretch}{1.24}    
\setlength{\jot}{6pt}       
\rnc{\arraystretch}{1.24}   

\makeatletter
\rnc{\theequation}{\thesection.\arabic{equation}}
\@addtoreset{equation}{section}
\makeatother



\nc{\be}{\begin{equation}}

\nc{\ee}{\end{equation}}

\nc{\bea}{\begin{eqnarray}}

\nc{\eea}{\end{eqnarray}}

\nc{\xx}{\nonumber\\}

\nc{\ct}{\cite}

\nc{\la}{\label}

\nc{\eq}[1]{(\ref{#1})}

\nc{\newcaption}[1]{\centerline{\parbox{6in}{\caption{#1}}}}

\nc{\fig}[3]{

\begin{figure}
\centerline{\epsfxsize=#1\epsfbox{#2.eps}}
\newcaption{#3. \label{#2}}
\end{figure}
}


\def\CA{{\cal A}}
\def\CC{{\cal C}}
\def\CD{{\cal D}}
\def\CE{{\cal E}}
\def\CF{{\cal F}}
\def\CG{{\cal G}}
\def\CH{{\cal H}}
\def\CK{{\cal K}}
\def\CL{{\cal L}}
\def\CM{{\cal M}}
\def\CN{{\cal N}}
\def\CO{{\cal O}}
\def\CP{{\cal P}}
\def\CR{{\cal R}}
\def\CS{{\cal S}}
\def\CU{{\cal U}}
\def\CV{{\cal V}}
\def\CW{{\cal W}}
\def\CY{{\cal Y}}
\def\CZ{{\cal Z}}


\def\IB{{\hbox{{\rm I}\kern-.2em\hbox{\rm B}}}}
\def\IC{\,\,{\hbox{{\rm I}\kern-.50em\hbox{\bf C}}}}
\def\ID{{\hbox{{\rm I}\kern-.2em\hbox{\rm D}}}}
\def\IF{{\hbox{{\rm I}\kern-.2em\hbox{\rm F}}}}
\def\IH{{\hbox{{\rm I}\kern-.2em\hbox{\rm H}}}}
\def\IN{{\hbox{{\rm I}\kern-.2em\hbox{\rm N}}}}
\def\IP{{\hbox{{\rm I}\kern-.2em\hbox{\rm P}}}}
\def\IR{{\hbox{{\rm I}\kern-.2em\hbox{\rm R}}}}
\def\IZ{{\hbox{{\rm Z}\kern-.4em\hbox{\rm Z}}}}


\def\a{\alpha}
\def\b{\beta}
\def\d{\delta}
\def\ep{\epsilon}
\def\ga{\gamma}
\def\k{\kappa}
\def\l{\lambda}
\def\s{\sigma}
\def\t{\theta}
\def\w{\omega}
\def\G{\Gamma}


\def\half{\frac{1}{2}}
\def\dint#1#2{\int\limits_{#1}^{#2}}
\def\goto{\rightarrow}
\def\para{\parallel}
\def\brac#1{\langle #1 \rangle}
\def\curl{\nabla\times}
\def\div{\nabla\cdot}
\def\p{\partial}


\def\Tr{{\rm Tr}\,}
\def\det{{\rm det}}


\def\vare{\varepsilon}
\def\zbar{\bar{z}}
\def\wbar{\bar{w}}
\def\what#1{\widehat{#1}}


\def\ad{\dot{a}}
\def\bd{\dot{b}}
\def\cd{\dot{c}}
\def\dd{\dot{d}}
\def\so{SO(4)}
\def\bfr{{\bf R}}
\def\bfc{{\bf C}}
\def\bfz{{\bf Z}}

\begin{titlepage}


\hfill\parbox{3.7cm} {{\tt arXiv:1709.04914}}

\vspace{15mm}

\begin{center}
{\Large \bf  Dark Energy and Dark Matter in Emergent Gravity}

\vspace{10mm}

Jungjai Lee ${}^{a}$\footnote{jjlee@daejin.ac.kr} and Hyun Seok Yang ${}^b$\footnote{hsyang@gist.ac.kr}
\\[10mm]

${}^a$ {\sl Division of Mathematics and Physics, Daejin University, Pocheon, Gyeonggi 11159, Korea}

${}^b$ {\sl Department of Physics and Photon Science, Gwangju Institute of Science and Technology, Gwangju 61005, Korea}

\end{center}

\thispagestyle{empty}

\vskip1cm


\centerline{\bf ABSTRACT}
\vskip 4mm
\noindent
Emergent gravity can be applied to a large $N$ matrix model by considering the vacuum of a noncommutative (NC) Coulomb branch
that satisfies the Heisenberg algebra.
Due to the fact that IR fluctuations in the NC Coulomb branch always pair with UV fluctuations,
this UV/IR mixing is extendable to a macroscopic scale.
These vacuum fluctuations in the NC Coulomb branch are described by a four-dimensional NC $U(1)$ gauge theory.
The order parameter for the vacuum fluctuations is given by random four-vectors that
have their own causal structure in the commutative limit unlike the conventional cosmological models based on a scalar field theory coupled to gravity. We show that their causal structure results in
the different nature of gravitational interactions so that space-like fluctuations give rise to the repulsive
gravitational force while time-like fluctuations generate the attractive gravitational force.
Given the fact that the fluctuations are random in nature and we live in a (3+1)-dimensional spacetime,
the ratio of the repulsive vs. attractive components ends up being 3:1 = 75:25,
which is interestingly consistent with the dark components of the current universe.
If we include ordinary matters acting as an attractive gravitational force,
the emergent gravity could more accurately explain the dark side of our universe.
This work is an expanded version of the conference proceedings \cite{EPJ}.
\\



\vspace{1cm}

\today

\end{titlepage}

\renewcommand{\thefootnote}{\arabic{footnote}}
\setcounter{footnote}{0}

\section{Introduction}

Dark energy and dark matter are a great mystery of the 20th century physics, which has not been resolved yet
within the paradigm of the contemporary physics.
In retrospect, the resolution of a great puzzle requires the upheaval of a radical new physics.
Recall how the blackbody radiation and the photoelectric effect had been resolved at the beginning of
the 20th century \cite{dirac}.
We know that these problems could not be solved by simply modifying the Newtonian dynamics
and the classical electrodynamics. A radical new paradigm, the so-called quantum mechanics, was necessary
to solve these problems. If dark energy and dark matter would be such a case, i.e., the 21st century version of
the blackbody radiation and the photoelectric effect, they would not be understood by simply modifying
the general relativity and the quantum field theory. Another novel paradigm, a.k.a. quantum gravity,
may be necessary to understand the nature of dark energy and dark matter \cite{jackng}.

The concept of emergent gravity and spacetime recently activated by the AdS/CFT correspondence
advocates that spacetime is not a fundamental entity existent from the beginning but an emergent property
from a primal monad such as matrices. The emergent spacetime is a new fundamental paradigm
that allows a background-independent formulation of quantum gravity and opens a new perspective to resolve
the notorious problems in theoretical physics such as the cosmological constant problem, hierarchy problem,
dark energy, and dark matter \cite{hsy-2016}. Moreover the emergent spacetime picture admits a background-independent
description of the inflationary universe which has a sufficiently elegant and explanatory power
to defend the integrity of physics against the multiverse hypothesis \cite{hsy-inflation,hsy-taiwan}.
We emphasize that noncommutative (NC) spacetime necessarily implies emergent spacetime if spacetime
at microscopic scales should be viewed as NC \cite{q-emg}. We will elaborate the emergent gravity from a large $N$
matrix model by considering a vacuum in the NC Coulomb branch satisfying the Heisenberg algebra
and argue that dark energy and dark matter may arise as a cosmic ouroboros of quantum gravity due to
the coherent vacuum structure of spacetime.

It was pointed out in \cite{hsy-jhep09,hsy-jpcs12} that the emergent gravity from NC $U(1)$ gauge fields
resolves the cosmological constant problem in a surprising way and explains the nature of dark energy
as arising from the UV/IR mixing of vacuum fluctuations over a coherent spacetime vacuum.
However the possibility of dark matter has been overlooked in this approach partially due to
the mainstream faction based on the particle model of dark matters. Recently there has been an
encouraging mood from observations that the notion about the particle nature of dark matter may crumble.
Moreover there are several suggestions to explain the dark matter based on
the quantum condensate of light scalar fields \cite{sin,whu,witten},
the emergent gravity from quantum entanglements \cite{verlinde,sabin2017,exp-test}
as well as the modified gravity \cite{milgrom,mond}.
A similar scheme to ours was recently presented
to solve the dark matter/energy problems by modifying general relativity to incorporate stringy gravity
at short distances \cite{dft-dm}. Since the theoretical structure of physics is always much bigger
and richer than we thought, we should not be stuck only to fashionable ideas.
For this reason, we want to explore the whole physical consequences derived from the emergent gravity
from large $N$ matrices and NC $U(1)$ gauge fields (see recent reviews, \cite{hsy-review,review})
and clarify how dark matter and dark energy can
be emergent from complex microphysical interactions in quantum gravity, like waves in sand dunes.

This paper is organized as follows. In section 2, we explain how one can get a higher-dimensional NC $U(1)$
gauge theory by considering a vacuum in the NC Coulomb branch of a zero-dimensional matrix model
and map the NC algebra defined by NC $U(1)$ gauge fields to gravitational variables
in higher dimensions \cite{hsy-2016,q-emg}. The whole procedure is summarized in Fig. 1.
In section 3, we identify the Einstein equations derived from the four-dimensional NC $U(1)$ gauge fields
and then take the analytic continuation to the Lorentzian signature to apply the emergent gravity to
our four-dimensional Universe \cite{hsy-jhep09,hsy-jpcs12}.
We find that the energy-momentum tensor deduced from the NC $U(1)$ gauge fields
generates the repulsive gravitational force as well as the attractive force which can be identified with
the dark energy and dark matter, respectively, with the ratio $75:25$. We discuss how the inclusion of
ordinary matters in this scheme can change the ratio closely to the observational data and
explain the galaxy rotation curve.
In section 4, we summarize our approach with the emphasis on the new results obtained in this paper
and discuss why emergent spacetime is a new fundamental paradigm for quantum gravity,
that opens a novel perspective to resolve the notorious problems in theoretical physics
such as the cosmological constant problem, hierarchy problem, dark matter, and dark energy \cite{hsy-essay}.
Appendix A clarifies the issue on the UV/IR mixing in the NC Coulomb branch.

\section{Emergent Spacetime from Matrices}

Let us start with a zero-dimensional matrix model with four $N \times N$ Hermitian matrices,
$\{\phi_a \in \mathcal{A}_N | a=1, \cdots, 4\}$, whose action is given by
\begin{equation} \label{matrix-action}
 S = - \frac{1}{4g^2} \mathrm{Tr} [\phi_a, \phi_b]^2.
\end{equation}
Note that the matrix action \eq{matrix-action} has the $U(N)$ gauge symmetry as well as a global automorphism
given by
\begin{equation}\label{poincare-auto}
    \phi_a \to \phi'_a = {\Lambda_a}^b \phi_b + c_a 
\end{equation}
if ${\Lambda_a}^b$ is a rotation in $SO(4)$ and $c_a$ are constants proportional
to the identity matrix. It will be shown later \cite{hsy-inflation} that the global symmetry \eq{poincare-auto}
is responsible for the Poincar\'e symmetry of flat spacetime emergent from a vacuum
in the Coulomb branch of matrix model and so will be called the Poincar\'e automorphism.

The equation of motion for the matrix model \eq{matrix-action} is given by
\begin{equation}\label{eom-mqm}
 [\phi_b, [\phi_a, \phi_b]] = 0,
\end{equation}
which must be supplemented with the Jacobi identity
\begin{equation}\label{gauss-mqm}
    [\phi_a, [\phi_b, \phi_c]] + [\phi_b, [\phi_c, \phi_a]] + [\phi_c, [\phi_a, \phi_b]] = 0.
\end{equation}
We want to study the dynamics of fluctuations around a vacuum in the Coulomb branch of the matrix model.
The conventional choice of vacuum in the Coulomb branch of $U(N)$ Yang-Mills theory is given by
$[\phi_a, \phi_b]|_{\mathrm{vac}} = 0 \; \Rightarrow \; \langle \phi_a \rangle_{\mathrm{vac}}
= \mathrm{diag}\big( (\alpha_a)_1, (\alpha_a)_2, \cdots, (\alpha_a)_N \big)$
for $a=1, \cdots, 4$. In this case the $U(N)$ gauge symmetry is broken to $U(1)^N$.
If we consider the $N \to \infty$ limit, the large $N$ limit opens a new phase of the Coulomb
branch given by
\begin{equation}\label{nc-coulomb}
    [\phi_a, \phi_b]|_{\mathrm{vac}} = - i B_{ab} \qquad \Rightarrow \qquad
    \langle \phi_a \rangle_{\mathrm{vac}} = p_a \equiv B_{ab} y^b
\end{equation}
where the vacuum moduli $y^a$ satisfy the Moyal-Heisenberg algebra $[y^a, y^b] = i \theta^{ab}$ with
$\theta^{ab} = (B^{-1})^{ab}$. This vacuum will be called the NC Coulomb branch.
Note that the NC Coulomb branch saves the NC nature of matrices while the conventional commutative vacuum
dismisses the property.

Suppose that the fluctuations around the vacuum \eq{nc-coulomb} take the form
\begin{equation}\label{gen-sol}
 \phi_a = p_a + \widehat{A}_a (y).
\end{equation}
We denote the NC $\star$-algebra on $\mathbb{R}^{4}_{\theta}$ by $\mathcal{A}_\theta$ and
$\widehat{A}_a (y) \in \mathcal{A}_\theta$ are four-dimensional NC $U(1)$ gauge fields.
The adjoint scalar fields in Eq. \eq{gen-sol} now obey the deformed algebra given by
\begin{equation}\label{def-moyal}
    [\phi_a, \phi_b] = - i (B_{ab} - \widehat{F}_{ab}) \in \mathcal{A}_\theta,
\end{equation}
where $\widehat{F}_{ab} = \partial_a \widehat{A}_b - \partial_b \widehat{A}_a - i [\widehat{A}_a, \widehat{A}_b]$
with the definition $\partial_a \equiv \mathrm{ad}_{p_a} = -i [p_a, \cdot]$.
Plugging the fluctuations in Eq. \eq{gen-sol} into the zero-dimensional matrix model (\ref{matrix-action}),
we get the four-dimensional NC $U(1)$ gauge theory. Thus we arrive at the following equivalence \cite{q-emg}:
\begin{equation} \label{rev-equiv}
 S = - \frac{1}{4g^2} \mathrm{Tr} [\phi_a, \phi_b]^2  = \frac{1}{4g_{YM}^2} \int d^4 y (\widehat{F}_{ab} - B_{ab})^2,
\end{equation}
where $g_{YM}^2 = (2\pi)^2 |\mathrm{Pf} \theta| g^2$ is the four-dimensional coupling constant.
It might be emphasized that the NC space \eq{nc-coulomb} is
a consistent vacuum solution of the action (\ref{matrix-action}) and the crux to realize
the equivalence \eq{rev-equiv}. If the conventional commutative vacuum were chosen,
we would have failed to realize the equivalence \eq{rev-equiv}.
Indeed it turns out \cite{hsy-2016,q-emg} that the NC Coulomb branch is crucial to realize the emergent gravity
from matrix models or large $N$ gauge theories.

We note that both $\mathcal{A}_N$ and $\mathcal{A}_\theta$ are associative
NC algebras and they form Lie algebras under their bracket operation.
In particular, the NC Coulomb branch \eq{nc-coulomb} admits a separable Hilbert space $\mathcal{H}$
and the NC field $\phi_a \in \mathcal{A}_\theta$ in Eq. \eq{gen-sol} can be identified with an element of
a linear map $\rho: \mathcal{H} \to \mathcal{H}$, i.e., $\rho = \mathrm{End}(\mathcal{H})$
and a linear representation $\rho: \mathcal{A}_\theta \to \mathcal{A}_N$ in $\mathcal{H}$ is
a Lie algebra homomorphism.
However there is another important lesson that we have learned from quantum mechanics \cite{dirac}.
For example, the momentum (position) operator in the Heisenberg algebra can be represented
by a differential operator in position (momentum) space, i.e., $p_i = -i\hbar \frac{\partial}{\partial x^i}$
or $x^i = i\hbar \frac{\partial}{\partial p_i}$.
More generally a NC algebra $\mathcal{A}_\theta$ has a representation in terms
of a differential (graded) Lie algebra $\mathfrak{D}$ and the map $\mathcal{A}_\theta \to \mathfrak{D}$
is also a Lie algebra homomorphism. To be specific, let us apply the Lie algebra homomorphism
$\mathcal{A}_\theta \to \mathfrak{D}$ to the dynamical variables in Eq. (\ref{gen-sol}).
We get a set of differential operators derived from the four-dimensional NC $U(1)$ gauge fields
on $\mathbb{R}_\theta^4$, which is defined by
\begin{equation}\label{nc-vector}
\widehat{V}_a = \{  \mathrm{ad}_{\phi_a} = -i [\phi_a, \, \cdot \,]| \phi_a (y) \in
\mathcal{A}_\theta \} \in \mathfrak{D}.
\end{equation}
In a large-distance limit, i.e. $|\theta| \to 0$, one can expand the NC vector fields $\widehat{V}_a$
using the explicit form of the Moyal $\star$-product.
The result takes the form
\begin{equation}\label{polyvector}
  \widehat{V}_a = V_a^\mu (y) \frac{\partial}{\partial y^\mu} + \sum_{p=2}^\infty
  V^{\mu_1 \cdots \mu_p}_a (y) \frac{\partial}{\partial y^{\mu_1}} \cdots
  \frac{\partial}{\partial y^{\mu_p}} \in \mathfrak{D},
\end{equation}
where $y^\mu$ are local coordinates on an emergent four-dimensional Riemannian manifold $M$.
Thus the expansion of NC vector fields in $\mathfrak{D}$ generates an infinite tower of
the so-called polyvector fields \cite{q-emg}.

\begin{figure}
\centering
\begin{picture}(400,250)


\put(116,242){\framebox[1.1\width]{0D $U(N \rightarrow \infty)$ Matrix model}}
\put(10,127){\framebox[1.1\width]{4D NC $U(1)$ gauge theory on $\mathbb{R}_\theta^4$}}
\put(260,127){\framebox[1.1\width]{4D Einstein gravity}}
\put(75,12){\framebox[1.1\width]{Differential operators as quantized frame bundle}}

\put(270,190){Large $N$ duality}
\thicklines
\textcolor[rgb]{0.00,0.00,1.00}{\put(180,230){\vector(-1,-1){85}}}

\put(30,190){NC Coulomb branch}

\thicklines
\textcolor[rgb]{0.98,0.00,0.00}{\put(220,230){\vector(1,-1){85}}}

\put(270,65){Classical limit}
\thicklines
\textcolor[rgb]{0.00,0.00,1.00}{\put(95,115){\vector(1,-1){85}}}

\put(55,65){Inner derivation}
\thicklines
\textcolor[rgb]{0.00,0.00,1.00}{\put(220,30){\vector(1,1){85}}}
\end{picture}
\caption{Emergent gravity is a large $N$ duality}
\label{fig-duality}
\end{figure}
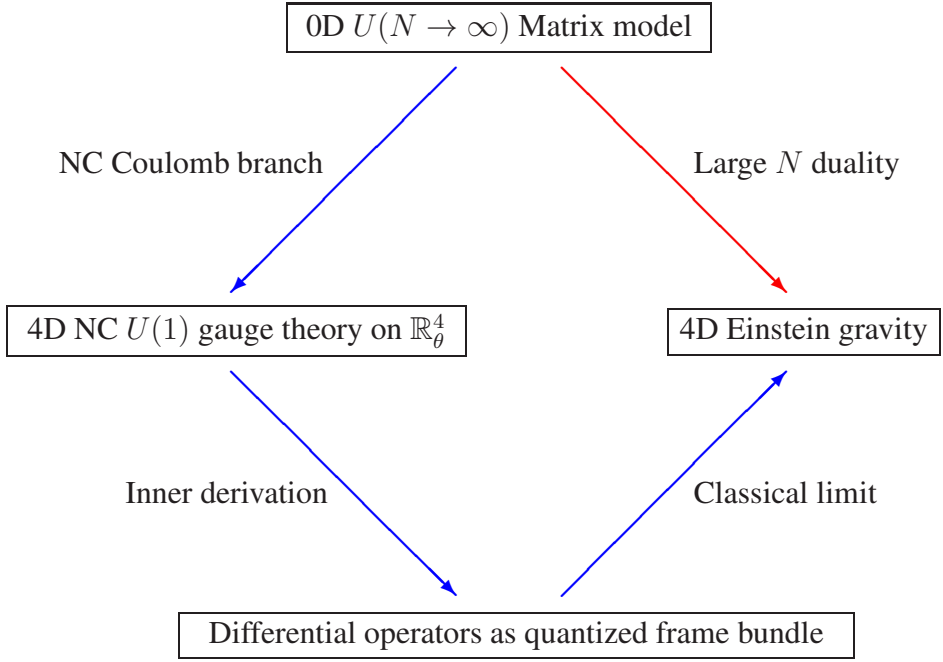

An interesting problem is to identify the theory described by the set of differential operators \eq{polyvector}.
Note that the leading term gives rise to the ordinary vector fields that will be identified
with a frame basis associated to the tangent bundle $TM$ of an emergent manifold $M$, as depicted in Fig. 1.
It is important to perceive that the realization of emergent geometry through the derivation algebra
in Eq. \eq{nc-vector} is intrinsically local \cite{q-emg}.
Therefore it is necessary to consider patching or gluing
together the local constructions to form a set of global quantities. We will assume that local
coordinate patches have been consistently glued together to yield global (poly)vector fields.
Let us denote the globally defined vector fields by
\begin{equation}\label{lorentzian-vec}
 \mathfrak{X}(M) = \Big\{ V_a = V_a^\mu (x) \frac{\partial}{\partial x^\mu}|
 a, \mu = 1, \cdots, 4 \Big\}.
\end{equation}
Define the structure equations of vector fields by
\begin{equation}\label{stv-eq}
    [V_a, V_b] = - {g_{ab}}^c V_c.
\end{equation}
The orthonormal vierbeins on $TM$ are then defined by the relation
$V_a = \lambda E_a \in \Gamma(TM)$ where the conformal factor $\lambda$ is determined
by $g_{bab} = V_a \ln \lambda^2$.
In the end, the Riemannian metric on a four-dimensional spacetime manifold $M$
is given by \cite{q-emg,hsy-jhep09}
\begin{equation}\label{eml-metric}
    ds^2 = g_{\mu\nu} dx^\mu dx^\nu = \lambda^2 v^a_\mu v^a_\nu dx^\mu dx^\nu
\end{equation}
where $v^a = v^a_\mu (x) dx^\mu$ are dual covectors defined by $V_a^\mu v^b_\mu = \delta^b_a$.

Let us trace the metric \eq{eml-metric} out to see where the flat spacetime comes from.
Definitely the flat spacetime corresponds to the vector field $V_a = \delta^\mu_a \frac{\partial}{\partial x^\mu}$
for which $\lambda^2 = 1$. It is easy to see that the flat spacetime arises from the vacuum
gauge fields $\langle \phi_a \rangle_{\mathrm{vac}} = p_a$ in Eq. \eq{nc-coulomb}.
This implies a remarkable picture \cite{hsy-jhep09,hsy-jpcs12}
that the flat spacetime is not an empty space unlike the general relativity
but emergent from a uniform vacuum condensate corresponding to a cosmological constant in general relativity.
In this case, the automorphism \eq{poincare-auto} precisely corresponds to the Poincar\'e symmetry of
flat spacetime. Therefore the NC Coulomb branch \eq{nc-coulomb} does not break the Poincar\'e symmetry.
Rather it is emergent from the symmetry \eq{poincare-auto} of the underlying matrix model.

\section{Dark Energy and Dark Matter from Emergent Gravity Picture}

In this section, we explore the physical consequences derived from the picture in Fig. 1.
First, note that $[\phi_a, [\phi_b, \phi_c]] = - \widehat{D}_a \widehat{F}_{bc}$. Using this result, one can derive
the equations of motion and the Bianchi identity of NC $U(1)$ gauge fields
from Eqs. \eq{eom-mqm} and \eq{gauss-mqm}, respectively. For these field variables,
the Lie algebra homomorphism reads as $\mathcal{A}_\theta \to \mathfrak{D}:
[\phi_a, [\phi_b, \phi_c]] \mapsto - [\widehat{V}_a, [\widehat{V}_b, \widehat{V}_c]]$ \cite{q-emg,hsy-jhep09}.
In the commutative limit, we thus get the following correspondence:
\begin{eqnarray} \label{cor-eom}
  &&  \widehat{D}_b \widehat{F}_{ab} = 0 \quad \xRightarrow{|\theta| \to 0} \quad
  [V_b, [V_a, V_b] ] = 0, \\
  \label{cor-bianchi}
  &&  \widehat{D}_{[a} \widehat{F}_{bc]} = 0 \quad \xRightarrow{|\theta| \to 0} \quad
  [V_{[a}, [V_b,V_{c]}]] = 0.
\end{eqnarray}
Note that the torsion and the curvature are multi-linear differential operators defined by
\begin{eqnarray} \label{def-tor}
&& T(X, Y) = \nabla_X Y - \nabla_Y X - [X, Y], \\
\label{def-cur}
&& R(X, Y) Z= [\nabla_X, \nabla_Y]Z -\nabla_{[X,Y]} Z,
\end{eqnarray}
where $X, Y$ and $Z$ are vector fields on $M$. Therefore they satisfy the relations
$T(V_a, V_b) = \lambda^2 T(E_a, E_b)$ and $R(V_a, V_b) V_c = \lambda^3 R(E_a, E_b) E_c$.
After imposing the torsion free condition, $T(E_a, E_b)=0$, it is easy to derive the identity
$R(E_{[a}, E_b) E_{c]} = \lambda^{-3} R(V_{[a}, V_b) V_{c]} = \lambda^{-3} [V_{[a}, [V_b,V_{c]}]] $.
Consequently we see \cite{hsy-jhep09,hsy-review} that the Bianchi identity \eq{cor-bianchi}
for NC $U(1)$ gauge fields in the commutative limit is equivalent to the first Bianchi identity
for the Riemann curvature tensors, i.e.,
\begin{equation}\label{bianchi-ncgr}
  \widehat{D}_{[a} \widehat{F}_{bc]} = 0 \quad \xRightarrow{|\theta| \to 0} \quad  R(E_{[a}, E_b) E_{c]} =0.
\end{equation}

The mission for the equations of motion \eq{cor-eom} is more involved. But, from the experience
on the Bianchi identity \eq{bianchi-ncgr}, we basically expect that it will be reduced to the Einstein equations
\begin{equation}\label{eom-ncgr}
  \widehat{D}_{b} \widehat{F}_{ab} = 0 \quad \xRightarrow{|\theta| \to 0} \quad
  R_{ab} = 8 \pi G \Big( T_{ab} - \frac{1}{2} \delta_{ab} T \Big).
\end{equation}
After a straightforward but tedious calculation, we get a remarkably simple
but cryptic result \cite{hsy-jhep09,hsy-review}
\begin{equation}\label{ricci-emt}
    R_{ab} = - \frac{1}{\lambda^2} \Big( g^{(+)i}_d g^{(-)j}_d (\eta^i_{ac} \overline{\eta}^j_{bc}
    + \eta^i_{bc} \overline{\eta}^j_{ac}) - g^{(+)i}_c g^{(-)j}_d (\eta^i_{ac} \overline{\eta}^j_{bd}
    + \eta^i_{bc} \overline{\eta}^j_{ad}) \Big).
\end{equation}
To get the above result, we have taken the canonical decomposition of the structure equation \eq{stv-eq} as
\begin{equation}\label{dec-ste}
  g_{abc} = g^{(+)i}_c \eta^i_{ab} + g^{(-)i}_c \overline{\eta}^i_{ab}.
\end{equation}

First it is convenient to decompose the energy-momentum tensor into two parts
\begin{eqnarray} \label{emt-max}
 &&  8 \pi G T_{ab}^{(M)} = - \frac{1}{\lambda^2} \Big(g_{acd} g_{bcd}
 - \frac{1}{4} \delta_{ab} g_{cde}g_{cde} \Big), \\
 \label{emt-lp}
  && 8 \pi G T_{ab}^{(L)} = \frac{1}{2 \lambda^2} \Big(\rho_{a} \rho_{b} - \Psi_a \Psi_b
 - \frac{1}{2} \delta_{ab} (\rho_{c} \rho_{c} - \Psi_c \Psi_c) \Big),
\end{eqnarray}
where $\rho_a \equiv g_{bab}$ and $\Psi_a \equiv - \frac{1}{2} \varepsilon^{abcd} g_{bcd}$.
A close inspection reveals that the first one is the Maxwell energy-momentum
tensor given by \cite{hsy-jhep09,hsy-jpcs12}
\begin{equation}\label{maxw-emt}
  T_{ab}^{(M)} = \frac{\hbar^2 c^2}{g_{YM}^2} \Big(F_{ac} F_{bc}
 - \frac{1}{4} \delta_{ab} F_{cd}F_{cd} \Big)
\end{equation}
but the second one seems to be mystic.

The reason is the following. The curvature tensor $R_{abcd}$ in general relativity can be decomposed
according to the canonical split of Lie algebra $so(4) = su(2)_L \oplus su(2)_R$ and the canonical
decomposition of two-forms $\Omega^2 (M) = \Omega^2_+ \oplus \Omega^2_-$ as \cite{ahs,riem-ym}
\begin{equation}\label{dec-riemann}
    R = \left(
          \begin{array}{cc}
            W^+ + \frac{1}{12} s & B \\
            B^T & W^- + \frac{1}{12} s \\
          \end{array}
        \right).
\end{equation}
A notable point is that the Ricci scalar $s$ appears in the blocks $(++)$ and $(--)$ while
the traceless Ricci tensors $B$ and $B^T$ show up in the blocks $(+-)$ and $(-+)$.
Note that the Ricci tensors \eq{ricci-emt} emergent from NC $U(1)$ gauge fields belong to the mixed sector
$(+-)$ and $(-+)$. Thus they should be traceless according the decomposition \eq{dec-riemann}
if they would be based on the general relativity.
However the Ricci tensor \eq{ricci-emt} has a nontrivial Ricci scalar given by
$s= \frac{1}{2 \lambda^2} (\rho_{a} \rho_{a} - \Psi_a \Psi_a)$.
Hence the Liouville energy-momentum tensor \eq{emt-lp} cannot be realized in the context of
general relativity. In order to descry more aspects of the second energy-momentum tensor,
let us separate the scalar and tensor perturbations as $\rho_{a} \rho_{b} =
\frac{1}{4} \delta_{ab} \rho_{c} \rho_{c} + \Big(\rho_{a} \rho_{b}
- \frac{1}{4} \delta_{ab} \rho_{c} \rho_{c} \Big)$ and
$\Psi_a \Psi_b = \frac{1}{4} \delta_{ab} \Psi_c \Psi_c + \Big(\Psi_{a} \Psi_{b}
- \frac{1}{4} \delta_{ab} \Psi_{c} \Psi_{c} \Big)$. In a long wavelength limit where
the quadruple modes can be ignored, the Liouville energy-momentum tensor \eq{emt-lp} behaves like
a cosmological constant
\begin{equation}\label{cc-r}
  T_{\mu\nu}^{(L)} \approx - \frac{s}{32\pi G} g_{\mu\nu}.
\end{equation}

In order to get a corresponding result in (3+1)-dimensional Lorentzian spacetime,
let us take the analytic continuation defined by $x^4 = ix^0$.
Under this Wick rotation, $g_{\mu\nu} \to g_{\mu\nu}, \; \rho_\mu \to \rho_\mu$ and $\Psi_\mu \to i \Psi_\mu$,
so the Liouville energy-momentum tensor in the Lorentzian signature is given by
\begin{equation}\label{lorentz-emt}
T_{\mu\nu}^{(L)} = \frac{1}{16 \pi G \lambda^2} \Big(\rho_{\mu} \rho_{\nu} + \Psi_\mu \Psi_\nu
 - \frac{1}{2} g_{\mu\nu} (\rho^2_{\lambda} + \Psi^2_\lambda) \Big).
\end{equation}
Note that $\rho_\mu$ and $\Psi_\mu$ are four-vectors and random fluctuations in nature.
Lorentzian four-vectors have their own causal structure unlike the Riemannian case.
They are classified into three classes:

(A) $(\rho_\mu, \Psi_\mu)$ are space-like vectors, i.e. $g^{\mu\nu} \rho_\mu \rho_\nu > 0$, etc.

(B) $(\rho_\mu, \Psi_\mu)$ are time-like vectors, i.e. $g^{\mu\nu} \rho_\mu \rho_\nu < 0$, etc.

(C) $(\rho_\mu, \Psi_\mu)$ are null vectors, i.e. $g^{\mu\nu} \rho_\mu \rho_\nu = 0$, etc. \\
which is in sharp contrast to the Riemannian space where every vectors are positive-definite.
We will see that their causal structure results in the different nature of gravitational interactions.

Let us clarify this important issue. Given a time-like unit vector $u_\mu$, i.e. $u^\mu u_\mu = -1$,
the Raychaudhuri equation in four dimensions is given by \cite{hawk-elli,emm-book}
\begin{equation}\label{ray-eq}
    \dot{\Theta} - \dot{u}^\mu_{;\mu} + \Sigma_{\mu\nu} \Sigma^{\mu\nu} - \Omega_{\mu\nu} \Omega^{\mu\nu}
    + \frac{1}{3}\Theta^2 = - R_{\mu\nu} u^\mu u^\nu,
\end{equation}
where $R_{\mu\nu} u^\mu u^\nu = \frac{1}{2 \lambda^2}  u^\mu u^\nu (\rho_{\mu} \rho_{\nu} + \Psi_\mu \Psi_\nu)$.
Assume, for simplicity, that all the terms except the expansion evolution, $\dot{\Theta}$,
in the Raychaudhuri equation vanish or become negligible. In this case the Raychaudhuri equation reduces to
\begin{equation}\label{ray-redeq}
    \dot{\Theta} = - \frac{1}{2 \lambda^2}  u^\mu u^\nu (\rho_{\mu} \rho_{\nu} + \Psi_\mu \Psi_\nu).
\end{equation}
In macroscopic scales where the scalar fluctuations are dominant compared to tensor perturbations,
the Raychaudhuri equation \eq{ray-redeq} can be approximated as
\begin{equation}\label{ray-maceq}
    \dot{\Theta} \approx  \frac{s}{4}
\end{equation}
where $s = \frac{1}{2 \lambda^2}  g^{\mu\nu} (\rho_{\mu} \rho_{\nu} + \Psi_\mu \Psi_\nu)$ is the Ricci scalar
determined by the Einstein equations \eq{eom-ncgr}.
Thus the behavior of spacetime expansion/contraction crucially depends on the causal structure of
random fluctuations in the three classes:

(A) $\dot{\Theta} \approx  \frac{s}{4} > 0$ when $(\rho_\mu, \Psi_\mu)$ are space-like vectors,

(B) $\dot{\Theta} \approx  \frac{s}{4} < 0$ when $(\rho_\mu, \Psi_\mu)$ are time-like vectors,

(C) $\dot{\Theta} \approx  \frac{s}{4} = 0$ when $(\rho_\mu, \Psi_\mu)$ are null vectors. \\
As a result, space-like fluctuations give rise to the repulsive gravitational force while
time-like fluctuations generate the attractive gravitational force.
Let us forget about the null case (C) since it does not contribute to the spacetime expansion/contraction.

In the above argument, we have assumed that the fluctuations $\rho_\mu$ and $\Psi_\mu$ share
the same causal structure since they all come from the structure equation \eq{stv-eq}
like the electric and magnetic fields in electromagnetism although we could not prove it rigorously.
The Liouville energy-momentum tensor \eq{lorentz-emt} stipulates an inimitable energy condition.
For a time-like unit vector $u^\mu$, it obeys the following property
\begin{equation}\label{wen-cond}
    T_{\mu\nu}^{(L)} u^\mu u^\nu = \frac{1}{32 \pi G \lambda^2} (g^{\mu\nu} + 2u^\mu u^\nu)
    (\rho_\mu \rho_\nu + \Psi_\mu \Psi_\nu) \geq 0
\end{equation}
since $g^{\mu\nu} + 2u^\mu u^\nu$ is a positive definite metric \cite{hawk-elli}
and
\begin{equation}\label{sen-cond}
    \Big( T_{\mu\nu}^{(L)} - \frac{1}{2} g_{\mu\nu} T^{(L)} \Big) u^\mu u^\nu
    = \frac{1}{16 \pi G \lambda^2} \Big( (\rho_\mu u^\mu)^2 + (\Psi_\mu u^\mu)^2 \Big) \geq 0.
\end{equation}
One can see that the equality in the above inequalities holds only for the Minkowski spacetime.
Remarkably the Liouville energy-momentum tensor \eq{lorentz-emt}
satisfies the weak energy condition \eq{wen-cond}
as well as the strong energy condition \eq{sen-cond} in spite of its exotic nature.
However, at large distances where shear modes can be ignored, it reduces to
\begin{equation}\label{laen-cond}
    T_{\mu\nu}^{(L)} u^\mu u^\nu = \frac{1}{64 \pi G \lambda^2} (\rho_\lambda^2 + \Psi_\lambda^2 )
    = - \Big( T_{\mu\nu}^{(L)} - \frac{1}{2} g_{\mu\nu} T^{(L)} \Big) u^\mu u^\nu.
\end{equation}
Therefore the cases (A) and (C) satisfy the weak energy condition while the case (B) does not if scalar modes
are large compared to tensor modes.
However the case (A) violates the strong energy condition whereas the cases (B) and (C) satisfy it
for the same situation. This means that the time-like fluctuations should not be interpreted
as ordinary matters and the space-like fluctuations behave like a cosmological constant or dark energy
at large distances. We will see that this observation is indeed true with interesting implications
for our Universe.

It should be interesting to estimate the energy scale of the Liouville energy-momentum tensor \eq{lorentz-emt}.
First observe that the NC Coulomb branch is generated by the Planck energy condensate in vacuum, i.e.,
$\rho_{vac} = \frac{1}{4g_{YM}^2} |B_{\mu\nu}|^2 \sim 10^{-2} M_P^4$ \cite{hsy-jhep09,hsy-jpcs12}.
An important point is that the fluctuations over the coherent vacuum \eq{nc-coulomb} should be subject
to the spacetime uncertainty relation since the NC Coulomb branch satisfies the Heisenberg algebra.
The spacetime uncertainty relation is realized as the UV/IR mixing of vacuum fluctuations \cite{uv-ir}.
See Appendix A for the discussion on the UV/IR mixing in the NC Coulomb branch.
This means that UV fluctuations in the NC Coulomb branch are always paired with IR fluctuations
and these UV/IR fluctuations can be extended to macroscopic scales \cite{hsy-jpcs12}.
Hence let us consider the vacuum fluctuations whose IR fluctuations have a macroscopic wavelength $L_H = M_H^{-1}$.
Having this in mind, let us estimate the energy density of the vacuum fluctuations.
Since the order estimate can be recast with the Euclidean action \eq{rev-equiv} too,
let us use it for simplicity. The energy density from Eq. \eq{rev-equiv} reads as
\begin{eqnarray}\label{energy-den}
    \rho = \rho_{vac} + \delta \rho &=& \frac{1}{4g_{YM}^2} \big(B_{\mu\nu}B^{\mu\nu}
    - 2 B_{\mu\nu} \widehat{F}^{\mu\nu} + \widehat{F}_{\mu\nu}\widehat{F}^{\mu\nu} \big) \xx
    &=& M_P^4 \Big( 1 + \frac{L_P^2}{L_H^2} \Big)^2 = M_P^4  + \frac{1}{L_P^2 L_H^2} + M_H^4,
\end{eqnarray}
where we used the fact $\rho_{vac} \simeq M_P^4$
and $|\widehat{F}_{\mu\nu}| = \frac{1}{L_H^2}$ for a simple dimensional reason.
It should be remarked that the cross term $B_{\mu\nu} \widehat{F}^{\mu\nu}$ is a total derivative term
but it cannot be dropped since the vacuum fluctuations represented by $\widehat{F}_{\mu\nu} (y)$ are
not localized but extended to a macroscopic scale $L_H$. Actually the above analysis shows that the vacuum
fluctuation energy
\begin{equation}\label{dark-energy}
  \delta \rho \sim  - \frac{1}{2g_{YM}^2} B_{\mu\nu} \widehat{F}^{\mu\nu} \approx \frac{1}{L_P^2 L_H^2}
\end{equation}
is coming from the boundary term on a hypersurface of radius $L_H$.
The same analysis shows \cite{hsy-jhep09} that the Ricci scalar in Eq. \eq{cc-r} has the curvature scale given
by $|s| \sim \frac{1}{L_H^2}$, i.e.,
\begin{equation}\label{dark-emt}
  T_{\mu\nu}^{(L)} \sim \pm \frac{1}{L_P^2 L_H^2} g_{\mu\nu}.
\end{equation}

A few remarks are in order.
If the macroscopic scale $L_H$ is identified with the size of cosmic horizon of our observable universe,
$L_H =1.3 \times 10^{26}m$, the extended (nonlocal) energy in Eq. \eq{dark-energy} or \eq{dark-emt} is
in good agreement with the observed value of current dark energy $\rho_{DE} := \delta \rho \approx (10^{-3} eV)^4$.
Moreover one can determine the total energy within the hypersurface of radius $L_H$,
which is given by $\delta E = \frac{4 \pi L_H}{3 L_P^2}$.
Thus the corresponding total entropy $\delta S = \frac{\delta E}{T_H}$ is determined as $\delta S = \frac{A_H}{4G}$
since the de Sitter temperature of the cosmological horizon is given by $T_H = \frac{1}{2 \pi L_H}$ \cite{gibb-hawk},
where $A_H = 4 \pi L_H^2$ and $8 \pi G = L_P^2$. Of course the numerical factor is a wishful thinking.
This argument shows that the dark energy/matter in our Universe would be a holographic manifestation
of a microscopic physics, a.k.a. quantum gravity.
We showed before that space-like fluctuations give rise to the repulsive gravitational force while
time-like fluctuations generate the attractive gravitational force.
When considering the fact that
the fluctuations are random in nature and we are living in the (3+1)-dimensional spacetime,
the ratio of the repulsive and attractive components will end in $\frac{3}{4}: \frac{1}{4}=75:25$
and this ratio curiously coincides with the dark composition of our current Universe \cite{hsy-jpcs12}.
Note that the dark energy in \eq{dark-emt} sets the current Hubble parameter $H_0 = \frac{c}{L_H}$
and the Hubble parameter induces a characteristic acceleration scale $a_0 = c H_0 = \frac{c^2}{L_H}$.
Since the dark matter is the third of the dark energy, the dark matter will give rise to the attractive
acceleration scale $\frac{a_0}{3} = \frac{c^2}{3 L_H}$.
This attractive force will compete with ordinary matters depending on their characteristic scales.
Since the ordinary matter acts as the attractive gravitational force too,
the inclusion of ordinary matters definitely changes the previous ratio as $75 {\downarrow}:25 {\uparrow}$,
that will cause a better match with the current observation.
Therefore it is expected that the emergent gravity can explain the dark sector of
our Universe more precisely after including ordinary matters in this scheme.

\section{Raychaudhuri equation with ordinary matters}

We will provide a more quantitative analysis on the effect of ordinary matters on the evolution
of our Universe in the presence of dark matter and dark energy.
We first analyze the effect of ordinary matters on the evolution of our Universe
in the presence of dark matter and dark energy.
For this purpose, let us simply add ordinary matters
in the background of dark matter and dark energy
described by the energy-momentum tensor \eq{lorentz-emt}.
The total energy-momentum tensor is then given by
\begin{equation}\label{total-emt}
    T_{\mu\nu} = T^{(L)}_{\mu\nu} + T^{(SM)}_{\mu\nu}
\end{equation}
where $T^{(SM)}_{\mu\nu}$ is the energy-momentum tensor for all fields in Standard Model including
the Maxwell energy-momentum tensor \eq{maxw-emt} (to be precise, its analytically continued version
to the Lorentzian signature). Similarly to Eq. \eq{ray-eq}, we consider the evolution equation
for the expansion $(\Theta)$, shear $(\Sigma^{\mu\nu})$ and rotation $(\Omega_{\mu\nu})$ along the flow
representing a time-like or null vector $u_\mu$, i.e. $u^\mu u_\mu = -1$ or $0$.
The Raychaudhuri equation in this case is given by \cite{hawk-elli,emm-book}
\begin{equation}\label{tray-eq}
    \dot{\Theta} + \Sigma_{\mu\nu} \Sigma^{\mu\nu} - \Omega_{\mu\nu} \Omega^{\mu\nu}
    + \frac{1}{n}\Theta^2 = - R_{\mu\nu} u^\mu u^\nu,
\end{equation}
where $n=3 \; (2)$ for the time-like (null) vector $u_\mu$.
Note that we have dropped an external force term $- \dot{u}^\mu_{;\mu}$ since we will consider only
geodesic congruences. For the energy-momentum tensor \eq{total-emt},
the right-hand side of Eq. \eq{tray-eq} is given by
\begin{equation}\label{ray-rhs}
    - R_{\mu\nu} u^\mu u^\nu = - \frac{1}{2 \lambda^2}  u^\mu u^\nu (\rho_{\mu} \rho_{\nu} + \Psi_\mu \Psi_\nu)
    - 8 \pi G \Big( T^{(SM)}_{\mu\nu} u^\mu u^\nu + \frac{(n-2)}{2} T^{(SM)} \Big).
\end{equation}

It is known \cite{emm-book} that all normal matters obey the strong energy condition.\footnote{The Higgs potential term,
$V(\phi) = \lambda (|\phi|^2 - \upsilon^2)^2$, in Standard Model can violate it,
but the Higgs field condenses to the vacuum after the electroweak epoch $\sim 10^{-32}$ s,
that is, $ \phi (x) = \big(\upsilon + H (x) \big) e^{i\alpha (x)}$.
Thus the potential energy becomes $V(\phi) = \lambda H^2 (H + 2 \upsilon)^2$ after the electroweak epoch,
so localized near the Higgs field $H(x)$.} This means that the second term on the right-hand side
of Eq. \eq{ray-rhs} is negative. Since the CMB observations show that our Universe, at least, beyond supercluster
scales ($\gtrsim$ 100 Mpc) can approximately be described by a homogeneous and isotropic spacetime,
the Universe can be described by the FLRW metric given by
\begin{equation}\label{flrw}
    ds^2 = -dt^2 + a(t)^2 h_{ij}(x) dx^i dx^j.
\end{equation}
In this case, the shear and rotation of spacetime can be ignored beyond the supercluster scales.
Over these scales, the energy-momentum tensor of ordinary matters in Eq. \eq{ray-rhs}
may be approximated by a perfect fluid with the stress tensor given by
\begin{equation}\label{emt-perfect}
    T_{\mu\nu}^{(PF)} = (\rho + p) u_\mu u_\nu + p g_{\mu\nu},
\end{equation}
where $\rho$ is the energy density of fluid and $p$ is its pressure.
For time-like congruences, the strong energy condition for the form \eq{emt-perfect} can be written
as $\rho + 3p \geq 0$. The same approximation can be applied to the first term on the right-hand side
of Eq. \eq{ray-rhs} which leads to the result
\begin{equation}\label{lemt-appr}
    - \frac{1}{2 \lambda^2}  u^\mu u^\nu (\rho_{\mu} \rho_{\nu} + \Psi_\mu \Psi_\nu) \approx
    \frac{1}{8 \lambda^2}  g^{\mu\nu} (\rho_{\mu} \rho_{\nu} + \Psi_\mu \Psi_\nu) \equiv \frac{s}{4}.
\end{equation}
Although we have maintained the same notation as \eq{ray-maceq}, $s$ is no longer a Ricci scalar
because the energy-momentum tensor \eq{total-emt} modifies the Ricci scalar.
Note that, for the FLRW metric \eq{flrw}, $\dot{\Theta} + \frac{1}{3}\Theta^2 = 3 \frac{\ddot{a}}{a}$.
Therefore the Raychaudhuri equation \eq{tray-eq} in the homogeneous and isotropic universe is given by
\begin{equation}\label{tray-hi}
    3 \frac{\ddot{a}}{a} \approx \frac{s}{4} - 8 \pi G (\rho + 3p).
\end{equation}

\begin{figure}[htb]
      \begin{center}
           \includegraphics[width=7cm]{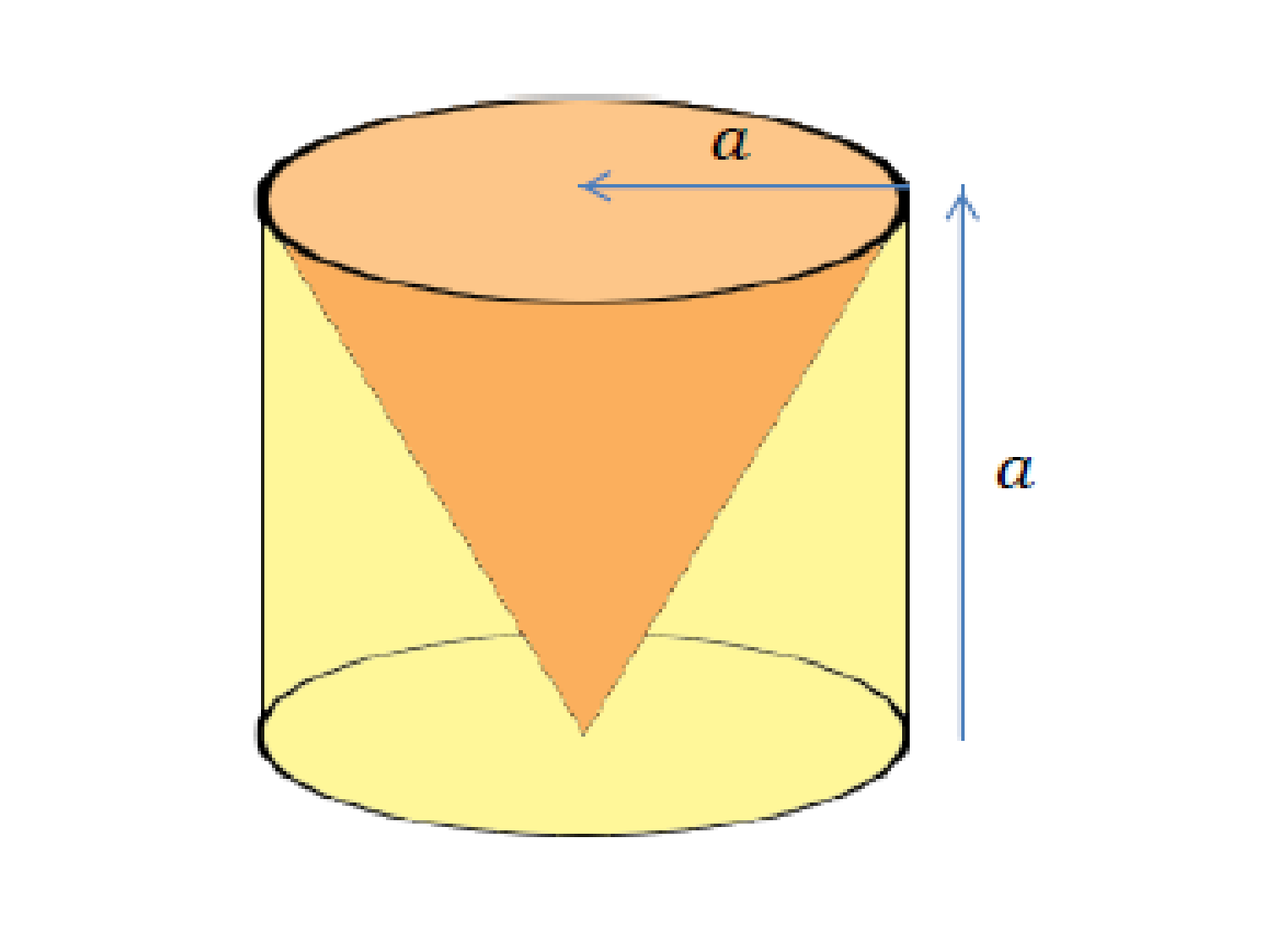}
       \end{center}
      \caption{Future light-cone inside a cylinder}
      \label{l-cone}
   \end{figure}

Since the causal structure of vacuum fluctuations cannot be mixed each other,
it should be considered separately:

(A) $3 \frac{\ddot{a}}{a}  = \frac{|s|}{4} - 8 \pi G (\rho + 3p) > 0$
when $(\rho_\mu, \Psi_\mu)$ are space-like vectors,

(B) $3 \frac{\ddot{a}}{a} = - \frac{|s|}{4} - 8 \pi G (\rho + 3p) < 0$
when $(\rho_\mu, \Psi_\mu)$ are time-like vectors,

(C) $3 \frac{\ddot{a}}{a} = - 8 \pi G (\rho + 3p) < 0$
when $(\rho_\mu, \Psi_\mu)$ are null vectors, \\
where, for the case (A), we used the fact that the dark energy is dominant 
over ordinary matters beyond the scale of galaxies.
As we observed before, space-like fluctuations in Eq. \eq{lemt-appr} give rise to the repulsive
gravitational force while time-like fluctuations generate the attractive gravitational force
with the ratio $\frac{3}{4}: \frac{1}{4}=75:25$ before including ordinary matters.
Let us clarify why the ratio should be obtained and the case (C) can be ignored compared to the cases (A) and (B).
Consider a cylinder with radius $a$ and height $a$ as a (2+1)-dimensional spacetime, for simplicity,
where the height signifies the time direction. Of course, we need to take the limit
$a \to \infty$ to describe macroscopic spacetimes. Consider a future-directed light-cone based on
the bottom inside the cylinder as shown in Fig. \ref{l-cone}. The time-like fluctuations should lie inside
the light-cone, so the available volume of them is $\frac{1}{3} \pi a^3$ while the space-like fluctuations
should lie outside the light-cone, so the available volume is $\frac{2}{3} \pi a^3$.
Hence the ratio of space-like vs. time-like fluctuations is $\frac{2}{3}: \frac{1}{3}$.
But the null fluctuations should be on the surface of light-cone, so its area is $\sqrt{2} \pi a^2$.
Thus the contribution from the null fluctuations can be ignored compared to the cases (A) and (B)
with the volume $\propto a^3$ when $a \to \infty$.
The same analysis can be generalized to a four-dimensional spacetime where
the ratio of space-like vs. time-like fluctuations is replaced by $\frac{3}{4}: \frac{1}{4}$ because
the volume of a $d$-dimensional cone is equal to $\frac{1}{d} \times (\mathrm{base}) \times (\mathrm{height})$.

Since $\rho + 3p > 0$ for ordinary matters,
the case (A) shows that ordinary matters tend to make the expansion rate reduce
while the case (B) shows that they tend to make the contraction rate increase.
This confirms our inference in section 3 that the inclusion of ordinary matters changes
the dark composition of our Universe as $75 {\downarrow}:25 {\uparrow}$.
In order to know a precise ratio, we need to know how much ordinary matters were produced through
quantum fluctuations during inflation and reheating, which is unfortunately beyond the reach yet.
Nevertheless, the current observational data indicate that the density of ordinary matter
in the universe is only 5 \% of the critical density;
the case (A) then suggests that dark energy proportion will be reduced to 70 \% = 75 \% - 5 \%.

Now let us consider the effect of dark matter on the rotation curve of galaxies.
Around the scale of galaxies ($\gtrsim$ kpc), the universe is neither homogeneous nor isotropic.
Thus the matter distribution of a galaxy would not be approximated by a perfect fluid.
Instead, the curvature deformation of spacetime by galaxies is small such that
post-Newtonian description in the background of dark matter can be applied to a good approximation.
In this approximation, the dark matter term \eq{lemt-appr} acts as an attractive force and
behaves like a cosmological constant $\frac{s}{12} = - \frac{1}{12 L_H^2} \equiv - \Lambda$
(anti-de Sitter space) because the shear (spin-2) modes in the post-Newtonian approximation
are generally small. In the presence of a cosmological constant, the Minkowski metric $\eta_{\mu\nu}$
is not a vacuum solution of the field equations, but for $\Lambda \ll 1$ an approximate solution
in a finite region can still be found by the expansion around $\eta_{\mu\nu}$ as
\begin{equation}\label{pn-metric}
    ds^2 = \Big(-1 + \frac{2\phi}{c^2} + \cdots \Big) dt^2
    + \delta_{ij} \Big(1 + \frac{2\phi}{c^2} + \cdots \Big) dx^i dx^j.
\end{equation}
The above metric is determined by directly solving Eq. \eq{ray-rhs} with $T_{\mu\nu}^{(SM)} \approx
\mathrm{diag} (\rho, 0, 0, 0) + \cdots$ where $\rho(\mathbf{x}, t)$ is a matter distribution of a galaxy.
For a more systematic expansion, see \cite{p-newton}.
To properly treat the $\Lambda$-term in the post-Newtonian approximation,
the size of the contributions due to $\Lambda$ must be at most comparable to the
post-Newtonian terms, which condition can be written as $\sqrt{\Lambda} \leq \frac{R_g}{r^2}$ where
$R_g = \frac{GM}{c^2}$ is the gravitational radius \cite{p-newton}. If $\sqrt{\Lambda} > \frac{R_g}{r^2}$,
the post-Newtonian approximation becomes bad and
the dark matter contribution becomes more important than ordinary matters. We will see that
this happens at a distance around the size of galaxies.

Finally we address the issue whether other epochs of the universe can be described equally well
by the energy-momentum tensor \eq{total-emt}. Let us consider the Einstein equations with
the stress tensor \eq{total-emt} for the FLRW metric \eq{flrw} whose $(00)$-component leads to
\begin{equation}\label{fried}
    3 H^2 = 8\pi G (\rho_{SM} + \rho_A + \rho_B + \rho_C),
\end{equation}
where $H(t) \equiv \dot{a}/a$ and $\rho_{SM} = T_{00}^{(SM)}$ and the energy density $T_{00}^{(L)}$ is divided
into three causal classes for space-like (A), time-like (B) and null (C) fluctuations.
The energy density $\rho_{SM}$ may be further decomposed into matter and radiation energy densities,
$\rho_M$ and $\rho_R$, so that $\rho_{SM} = \rho_M + \rho_R$. The expansion rate of the universe can
then be written as the well-known form \cite{emm-book}
\begin{equation}\label{ex-rate}
    H^2 = H_0^2 \big( \Omega_M (1 + z)^3 + \Omega_R (1 + z)^4 + \Omega_A (1 + z)^{3 (1 + \omega_A)}
    + \Omega_B (1 + z)^{3 (1 + \omega_B)}+ \Omega_C (1 + z)^{3 (1 + \omega_C)} \big),
\end{equation}
where $H_0$ is the present Hubble constant and $z$ is the redshift factor. Here we used the relation
$\Omega_i = \frac{\rho_{i,0}}{\rho_{crit}}$ and $\frac{\rho_i}{\rho_{i,0}} = (1 + z)^{3 (1 + \omega_i)}$
for species $i$ with the equation of state $\omega_i$ where $\rho_{crit} = \frac{3 H_0^2}{8 \pi G}$.
In order to know the expansion rate of the universe at time $t$, it is necessary to know the equation
of states $\omega_i$ besides $\Omega_i$ for $i = A, B,C$. They are not known to us so far and more works
are necessary to pin down their precise values. Hence let us take the approximation \eq{lemt-appr}.
In this limit, $T_{\mu\nu}^{(L)} \sim - \frac{s}{32 \pi G} g_{\mu\nu}$, so behaves like
a cosmological constant\footnote{In general, it is true only in a finite time period.
Since $\rho_a = g_{bab} = V_a \ln \lambda^2$ and $\lambda^2$ will be time-dependent in an expanding universe,
the Ricci scalar $s$ will be time-dependent too.} and obeys $\Omega_B = - \frac{1}{3} \Omega_A, \; \Omega_C = 0$
and $\omega_A = \omega_B \approx - 1$. Then Eq. \eq{ex-rate} reduces to
\begin{equation} \label{rex-rate}
    H^2 = H_0^2 \big( \Omega_M (1 + z)^3 + \Omega_R (1 + z)^4 + \frac{2}{3} \Omega_A \big).
\end{equation}
With the input, $\Omega_M = 0.05, \; \Omega_R = 8 \times 10^{-5}$ and $\Omega_A = 0.7$, Eq. \eq{rex-rate}
estimates the dark energy dominated era at $z \lesssim 1.1$. According to observational data,
the dark energy dominated era started at $z \lesssim 0.5$. Therefore, to understand an early Universe
at $z \gtrsim 0.5 \sim 1$, it is required to perform a more refined analysis by improving our crude approximation
taking a simple truncation to scalar modes and regarding the Ricci scalar as a cosmological constant.
Indeed, Eq. \eq{sen-cond} shows that the energy-momentum tensor $T_{\mu\nu}^{(L)}$ generates an attractive
force for any fluctuations if the shear modes are not ignored and causes the delay
of the dark energy dominated era.

The emergent gravity picture requires to unify geometry and matters on an equal footing.
But it is not yet understood what matter is from the emergent gravity picture
although a tough idea was suggested in \cite{q-emg,hsy-jhep09,hsy-jpcs12}.
So we simply assumed the matter fields that obey the law of Standard Model.
Suppose that ordinary matters are added in the background of dark matter and dark energy given by \eq{dark-emt}.
Then there are two independent sources generating the attractive force and
gravitational forces due to the dark matter and ordinary matters will compete each other
with their own characteristic scales. Let us only highlight the main argument.
Since the gravitational force generated by ordinary matters decays as the $\frac{1}{r^2}$-law,
the gravitational force due to ordinary matters will dominate at small scales while
the gravitational acceleration due to dark matter dominates at large scales.
To illuminate this aspect, let us consider a (spiral or disk) galaxy and $M(r)$ be the mass
contained inside an orbit of radius $r$.
The mass distribution of the galaxy gives rise to the acceleration $a_M = \frac{G M(r)}{r^2}$ which decreases
as $1/r^2$ as $r$ increases if there is no mass outside this radius. Thus there is a crossover where
the acceleration $a_M$ becomes equal to the acceleration $\frac{a_0}{3} = \frac{c^2}{3 L_H}$ due to the dark matter:
\begin{equation}\label{cross}
  a_M = \frac{G M}{r^2}  \lesssim \frac{a_0}{3} = \frac{c^2}{3 L_H}.
\end{equation}
One can see from Eq. \eq{cross} that this crossover arises at the distance $r_c = \frac{\sqrt{3 GM L_H}}{c}$
from the center of the galaxy, over which the acceleration $a_0/3$ due to the dark matter dominates.
For example, the crossover distances
for M33 $(M = 5 \times 10^{10} M_{\odot}, \; R = 9 \, \mathrm{kpc})$ and
the Milky Way $(M = 10^{12} M_{\odot}, \; R = 38 \, \mathrm{kpc})$
are $r_c \approx 5.6 \, \mathrm{kpc} \; (1.7 \times 10^{20} m)$ and $29.8 \, \mathrm{kpc} \;
(9.2 \times 10^{20} m)$, respectively,
which are roughly the size of the galaxies. This implies that the flattening of
the galaxy rotation curve may be explained by the dark matter given by Eq. \eq{dark-emt}.

\section{Discussion}

We have applied the emergent gravity to a large $N$ matrix model by considering a vacuum
in the NC Coulomb branch satisfying the Heisenberg algebra.
The vacuum fluctuations in the NC Coulomb branch are described by a four-dimensional NC $U(1)$ gauge theory
and thus the random four-vectors have their own causal structures in the commutative limit
unlike the conventional cosmological models based on a scalar field theory coupled to gravity.
We showed that their causal structure results in the different nature of gravitational interactions
so that space-like fluctuations give rise to the repulsive gravitational force while
time-like fluctuations generate the attractive gravitational force. When considering the fact that
the fluctuations are random in nature and we are living in the (3+1)-dimensional spacetime,
the ratio of the repulsive and attractive components ends in $\frac{3}{4}: \frac{1}{4}=75:25$.
We have performed a quantitative analysis to indicate that the inclusion of ordinary matters
in the background of dark matter and dark energy can explain the dark sector of our Universe more precisely.
Moreover we have illustrated how the existence of two attractive forces due to ordinary matter and dark matter
with completely different characteristic scales can explain the flattening of the galaxy rotation curve.
However, in order to understand the history of our Universe at $z \gtrsim 0.5 \sim 1$
from the emergent gravity picture, it seems to be necessary to consider a more sophisticated analysis
beyond taking a simple truncation to scalar modes and regarding the Ricci scalar as a cosmological constant.

The most remarkable aspect of emergent gravity is that any kind of spacetime structures should not
be assumed in advance but must be defined as a solution of an underlying background-independent theory.
Even the flat Minkowski spacetime should have its own dynamical origin defined by matrices
or NC $U(1)$ gauge fields \cite{hsy-2016}. We observed at the end of section 2 that the flat space $\mathbb{R}^4$
arises from the vacuum \eq{nc-coulomb} in the NC Coulomb branch which becomes the Minkowski spacetime
after the analytic continuation. And we have seen in Eq. \eq{energy-den} that the vacuum condensate
in the NC Coulomb branch naturally results from the dynamical condensate of the Planck energy.
The curved spacetime is generated by the deformations of the Coulomb branch vacuum
schematically represented by $\mathcal{F}_{\mu\nu} = B_{\mu\nu} + F_{\mu\nu} \; \Leftrightarrow \;
g_{\mu\nu} = \eta_{\mu\nu} + h_{\mu\nu}$ \cite{hsy-mirror}.
If the flat spacetime emerges from the Planck energy condensation in vacuum,
it implies that spacetime behaves like an elastic body with the tension of Planck energy.
In other words, gravitational fields generated by the deformations of the vacuum \eq{nc-coulomb}
will be very weak because the spacetime vacuum corresponds to a harmonic oscillator
with a stiffness of the Planck energy.
Therefore the dynamical origin of flat spacetime explains the metrical elasticity opposing
the curving of space and the stunning weakness of gravitational force \cite{hsy-jpcs12}.
Furthermore the emergent spacetime implies that the global Lorentz symmetry, being the isometry of flat spacetime,
should be a perfect symmetry up to the Planck scale because the flat spacetime was originated
from the condensation of the maximum energy in Nature.

The above spacetime picture in emergent gravity may bear some analogy with water waves in a swimming pool.
Without water in the swimming pool, it is not possible to generate the water wave but instead
sound waves can occur through air molecules in the pool.
In order to generate the water wave, first it is necessary to fill up the swimming pool with water.
Similarly there have been two phases of vacua in the Coulomb branch of a large $N$ matrix model:
the commutative vacuum and the NC vacuum. Unfortunately general relativity has no explanation about
the dynamical origin of the Minkowski spacetime and there is a tangible difference about the origin
of flat spacetime between general relativity and emergent gravity: the water in the swimming pool is regarded as
a completely empty space in general relativity. This misconception for the dynamical origin of spacetime
introduces several notorious puzzles in theoretical physics such as the cosmological constant problem
and the hierarchy problem \cite{hsy-jpcs12}.
In particular, the correct identification of the dynamical origin of flat spacetime
has been crucial to understand why dark energy and dark matter correspond to a cosmic ouroboros of quantum gravity
due to the coherent vacuum structure of spacetime. A more meditation about emergent spacetime also reveals
a remarkable picture \cite{hsy-inflation,hsy-taiwan} that the cosmic inflation corresponds to the dynamical emergence of
spacetime describing the dynamical process of Planck energy condensation in vacuum, i.e.,
the instant filling up the swimming pool with water.

\section*{Acknowledgments}

Some part of this work has been discussed in several international/domestic workshops: International Conference on Gravitation : Joint Conference of ICGAC-XIII and IK15 in July, 2017, 5th IBS Brainstorm Workshop in July, 2017 and International Workshop for String Theory and Cosmology in August, 2017.
We thank the organizers of each workshop for giving us an invaluable opportunity to present our work
and the participants for exchanging ideas and feedback.
This work was supported by the National Research Foundation of Korea (NRF) grants funded
by the Korea government (MSIT) (No. NRF-2020R1F1A1068410)(JL) and NRF-2018R1D1A1B0705011314 (HSY).


\appendix

\section{UV/IR mixing in the NC Coulomb Branch}

The UV/IR mixing is one of cruxes to understand the nature of dark energy and dark matter from
the emergent spacetime picture. In order to see why the NC Coulomb branch \eq{nc-coulomb}
satisfying the Heisenberg algebra necessarily gives rise to the UV/IR mixing,
consider the NC space $\mathbb{R}_\theta^{2n}$ whose coordinate generators satisfy the commutation relation
\begin{equation}\label{nc-r2n}
    [y^a, y^b ]= i\theta^{ab}, \qquad       (a,b=1,\cdots,2n),
\end{equation}
where $\theta^{ab}$ is a $2n \times 2n$ symplectic matrix. To be specific, let us take $\theta^{ab}$ as the form
\begin{equation} \label{symp-theta}
    \theta^{ab} = \alpha' \left(
                            \begin{array}{cc}
                              0 & \mathbb{I} \\
                              - \mathbb{I} & 0 \\
                            \end{array}
                          \right)
\end{equation}
where $\alpha'\equiv l_s^2$ is a fundamental constant with the physical dimension of $(\mathrm{length})^2$
and $\mathbb{I}$ is the $n \times n$ identity matrix.
Given a polarization like \eq{symp-theta}, it is convenient to split the coordinate generators
as $y^a = (y^i, y^{n+i}), \;  i=1,\cdots,n$, and rename them as $y^i \equiv x^i$ and
$y^{n+i} \equiv \frac{\alpha'}{\hbar} p^i$.
We have intentionally introduced the Planck constant $\hbar$.
Note that $[y^{n+i}]$ carries the physical dimension of length, as it should be, if $p^i$ is a momentum.
Then the commutation relation \eq{nc-r2n} can be written as
\begin{equation}\label{nc-phase}
[ x^i, p^j ]= i \hbar\delta^{ij}.
\end{equation}
Therefore we can apply the Heisenberg uncertainty principle to the commutation relation \eq{nc-phase}
which leads to
\begin{equation}\label{uncertain-qm}
    \Delta x^i  \Delta p^j  \geq \frac{\hbar}{2} \delta^{ij}.
\end{equation}
If we use the original variables $y^a = (y^i, y^{n+i})$, the above uncertainty relation reads as
\begin{equation}\label{uncertain-nc}
    \Delta y^i  \Delta y^{n+j}  \geq \frac{\alpha'}{2} \delta^{ij}.
\end{equation}

The commutation relation \eq{nc-phase} implies that the mathematical structure of NC space is the same
as quantum mechanics. Thus one can regard the physics on the NC space \eq{nc-r2n} as a `quantum mechanics'
defined by $\alpha'$ instead of $\hbar$. This is the reason why one should not consider
the NC space $\mathbb{R}_\theta^{2n}$
as a {\it classical} space. Moreover the scales where the noncommutative (or quantum) effect becomes significant
are dramatically different for the NC space and quantum mechanics. Since the noncommutativity of spacetime is set
by the fundamental constant $\alpha'\equiv l_s^2$, it is natural to consider the length scale as the Planck length,
i.e. $l_s= 10^{-35}$m. $l_s$ is much more smaller than the scale for quantum mechanics,
typically the Bohr radius $r_B = 5.3 \times 10^{-11}$m.
The Bohr radius corresponds to the size of superclusters ($\sim 10^{24}$m) in our Universe to an observer
who appreciates the noncommutative effect \eq{uncertain-nc} resulting from the NC space $\mathbb{R}_\theta^{2n}$.
Hence the quantum mechanics represented by the NC phase space \eq{nc-phase} rather behaves like a ``classical system"
to an observer near the Planck scale, so it may be natural to consider it as emergent
from something deeper \cite{witten-se}, possibly from the NC space \eq{nc-r2n}.
Therefore it is necessary to take the uncertainty relation \eq{uncertain-nc} into account as a primary effect
when we consider the physics on the NC space \eq{nc-r2n}.
Although the UV/IR mixing was derived in \cite{uv-ir} from quantum loops controlled by $\hbar$,
it is obvious from the uncertainty relation \eq{uncertain-nc} that the UV/IR mixing should exist even without
considering quantum mechanics, i.e. $\hbar$-effects, since the NC space \eq{nc-r2n} can be written
as the form \eq{nc-phase}.


\end{document}